\documentstyle[11pt,moriond,epsfig]{article}

\def\Journal#1#2#3#4{{#1} {\bf #2}, #3 (#4)}

\def\NPB{{\em Nucl. Phys.} B}
\def\PLB{{\em Phys. Lett.}  B}

\def\ppg{\pi^{+}\pi^{-}\gamma}
\def\mmg{\mu^{+}\mu^{-}\gamma}
\def\eeg{e^{+}e^{-}\gamma}

\def\be{\begin{equation}}
\def\ee{\end{equation}}
\def\bea{\begin{eqnarray}}
\def\eea{\end{eqnarray}}

\begin{document}
\vspace*{4cm}
\title{MEASUREMENT OF HADRONIC CROSS SECTION AT KLOE}

\author{The KLOE Collaboration~\footnote{
The KLOE Collaboration: A.~Aloisio,
F.~Ambrosino,
A.~Antonelli,
M.~Antonelli,
C.~Bacci,
G.~Bencivenni,
S.~Bertolucci,
C.~Bini,
C.~Bloise,
V.~Bocci,
F.~Bossi,
P.~Branchini,
S.~A.~Bulychjov,
R.~Caloi,
P.~Campana,
G.~Capon,
G.~Carboni,
M.~Casarsa,
V.~Casavola,
G.~Cataldi,
F.~Ceradini,
F.~Cervelli,
F.~Cevenini,
G.~Chiefari,
P.~Ciambrone,
S.~Conetti,
E.~De~Lucia,
G.~De~Robertis,
P.~De~Simone,
G.~De~Zorzi,
S.~Dell'Agnello,
A.~Denig,
A.~Di~Domenico,
C.~Di~Donato,
S.~Di~Falco,
A.~Doria,
M.~Dreucci,
O.~Erriquez,
A.~Farilla,
G.~Felici,
A.~Ferrari,
M.~L.~Ferrer,
G.~Finocchiaro,
C.~Forti,
A.~Franceschi,
P.~Franzini,
C.~Gatti,
P.~Gauzzi,
S.~Giovannella,
E.~Gorini,
F.~Grancagnolo,
E.~Graziani,
S.~W.~Han,
M.~Incagli,
L.~Ingrosso,
W.~Kluge,
C.~Kuo,
V.~Kulikov,
F.~Lacava,
G.~Lanfranchi,
J.~Lee-Franzini,
D.~Leone,
F.~Lu,
M.~Martemianov,
M.~Matsyuk,
W.~Mei,
L.~Merola,
R.~Messi,
S.~Miscetti,
M.~Moulson,
S.~M\"uller,
F.~Murtas,
M.~Napolitano,
A.~Nedosekin,
F.~Nguyen,
M.~Palutan,
L.~Paoluzi,
E.~Pasqualucci,
L.~Passalacqua,
A.~Passeri,
V.~Patera,
E.~Petrolo,
L.~Pontecorvo,
M.~Primavera,
F.~Ruggieri,
P.~Santangelo,
E.~Santovetti,
G.~Saracino,
R.~D.~Schamberger,
B.~Sciascia,
A.~Sciubba,
F.~Scuri,
I.~Sfiligoi,
T.~Spadaro,
E.~Spiriti,
G.~L.~Tong,
L.~Tortora,
E.~Valente,
P.~Valente,
B.~Valeriani,
G.~Venanzoni,
S.~Veneziano,
A.~Ventura,
G.~Xu,
G.~W.~Yu.
} 
\\presented by B. Valeriani}
\address{IEKP-Universit\"at Karslruhe, Postfach 3640, 76021, Karlsruhe, Germany}

\maketitle
\abstracts{In the fixed energy environment of the $e^{+}e^{-}$ collider DA$\Phi$NE, KLOE can measure the cross section of the process $e^{+}e^{-} \rightarrow$ hadrons as a function of the hadronic system energy using the radiative return method. At energies below 1 GeV, $e^{+}e^{-} \rightarrow \rho \rightarrow \pi^{+}\pi^{-}$ is the dominating hadronic process. We report here on the status of the analysis of the process $e^{+}e^{-} \rightarrow \ppg$. Some preliminary results on the invariant mass spectrum of the two pion system, obtained from the analysis of $\sim 22.6$ pb$^{-1}$, are presented.}
\section{Measuring hadronic cross section at DA$\Phi$NE}
\subsection{The radiative return method}\label{subsec:method}
DA$\Phi$NE~\cite{dafne} is an $e^{+}e^{-}$ storage-ring collider working at the $\phi$ resonance (1020 MeV). As an experiment at a collider with a fixed centre of mass energy, KLOE can measure the hadronic cross section $\sigma$($e^{+}e^{-} \rightarrow$ hadrons) as a function of the hadronic system energy using the {\em radiative return method}~\cite{spagnolo} ~\cite{bkm}, i. e. studying the process $e^{+} e^{-} \rightarrow$ hadrons$+\gamma$. The emission of one photon before the beams interact (Initial State Radiation) lowers the interaction energy and  makes possible to produce the hadronic system with an invariant mass varying from the $\phi$ mass down to the production threshold. 

The method represents an alternative approach to the conventional energy scan used so far for hadronic cross section measurements. A very solid theoretical understanding of Initial State Radiation (described by the radiation function $H$) is mandatory in order to extract the cross section $\sigma$($e^{+}e^{-} \rightarrow$ hadrons) as a function of $M_{had}^{2}$ from the measured differential cross section $d\sigma$($e^{+}e^{-} \rightarrow$ hadrons$+\gamma$)$/dM_{had}^{2}$:
\begin{equation}
M_{had}^{2} \cdot \frac{d\sigma(e^{+}e^{-} \rightarrow {\rm hadrons} + \gamma)}{dM_{had}^{2}} = \sigma(e^{+}e^{-} \rightarrow {\rm hadrons}) \cdot H(M_{had}^{2},\theta_{had})
\label{eq:radiation}
\end{equation}
where $M_{had}^{2}$ and $\theta_{had}$ are the invariant mass squared and the acceptance cut on the polar angle of the hadronic system. Radiative corrections to the Born process $e^{+}e^{-} \rightarrow$ hadrons have been calculated by different theoretical groups~\cite{rodrigo}~\cite{jege}~\cite{pancheri} up to the NLO for $e^{+}e^{-} \rightarrow \pi^{+}\pi^{-}$. The radiation function $H$ is known at present with an accuracy of better than $1\%$.  We want to stress that the radiative return method has the merit compared with the conventional energy scan that the systematics of the measurement (e.g. normalization, beam energy) are the same for any experimental point and must not be evaluated at each energy step.
\subsection{Hadronic cross section at KLOE}\label{subsec:motivi}
At energies below the $\phi$ mass, the main contribution to hadron production comes from the $\rho$ meson, i. e. from the process $e^{+}e^{-} \rightarrow \rho \rightarrow \pi^{+} \pi^{-}$. For this reason we concentrated at KLOE on the analysis of the radiative process $e^{+}e^{-} \rightarrow \ppg$. 

The measurement of hadronic cross section plays an important role in the reduction of the theoretical error of the muon anomalous magnetic moment, $a_{\mu}$, and of the running electromagnetic constant at the $Z$ mass, $\alpha(M_{Z})$, dominated by the uncertainty on the contribution of hadrons to the photon vacuum polarization. 
At low energies, where perturbative QCD is not applicable, the hadronic contributions to $a_{\mu}$ and to $\alpha(M_{Z})$ are evaluated via a dispersion integral using the measurements of hadronic cross sections. 
In the $e^{+}e^{-}$ data approach~\cite{jege2}, based on available $e^{+}e^{-} \rightarrow$ hadrons data up to 12 GeV, the energy range covered by KLOE is responsible for $\sim 67\%$ of the present error on $a_{\mu}$ and for $\sim 17\%$ of the error on $\alpha(M_{Z})$. The use of $\tau$ data in these calculations~\cite{davier} reduces the weight of the energy region below about 2 GeV but the validity of this method is not universally accepted. 
\section{Analysis of the process $e^{+}e^{-} \rightarrow \ppg$}
\subsection{Signal selection}\label{subsec:sig_sel}
The selection of $e^{+}e^{-} \rightarrow \ppg$ events is done in the following steps:
\begin{itemize} 
\item {\it detection of two charged tracks}, with polar angle larger than $40^o$, coming from a vertex in the fiducial volume $R<8$ cm, $|z|< 15$ cm. The cuts on the transverse momentum $p_{T} > 200$ MeV or on the longitudinal momentum $|p_{z}| > 90$ MeV reject tracks spiralizing along the beam line, ensuring good reconstruction conditions. 
The probability to reconstruct a vertex in the drift chamber is $\sim 95\%$ and has been studied with Bhabha data, selected using the calorimeter only. This efficiency and the track reconstruction efficiency are going to be also evaluated using $\pi^{+}\pi^{-}\pi^{0}$ events selected by detecting $\pi^{0}$ in the electromagnetic calorimeter. 
\item {\it identification of pion tracks}. A Likelihood Method (calibrated on real data), using the time of flight of the particle and the shape of the energy deposit in the electromagnetic calorimeter, has been developed to reject the $e^{+}e^{-} \rightarrow \eeg$ background. $\eeg$ events are drastically reduced. A control sample of $\pi^{+}\pi^{-}\pi^{0}$ has been used to study the behaviour of pions in the electromagnetic calorimeter and to evaluate the selection efficiency for signal events, which turns out to be larger than $98\%$. The effect of the selection on $\ppg$ and $\eeg$ events is visible in the left plot of Fig.~\ref{fig:trkmass}, as a function of the track mass, $M_{track}$. 
\begin{figure} 
\begin{center}
\psfig{figure=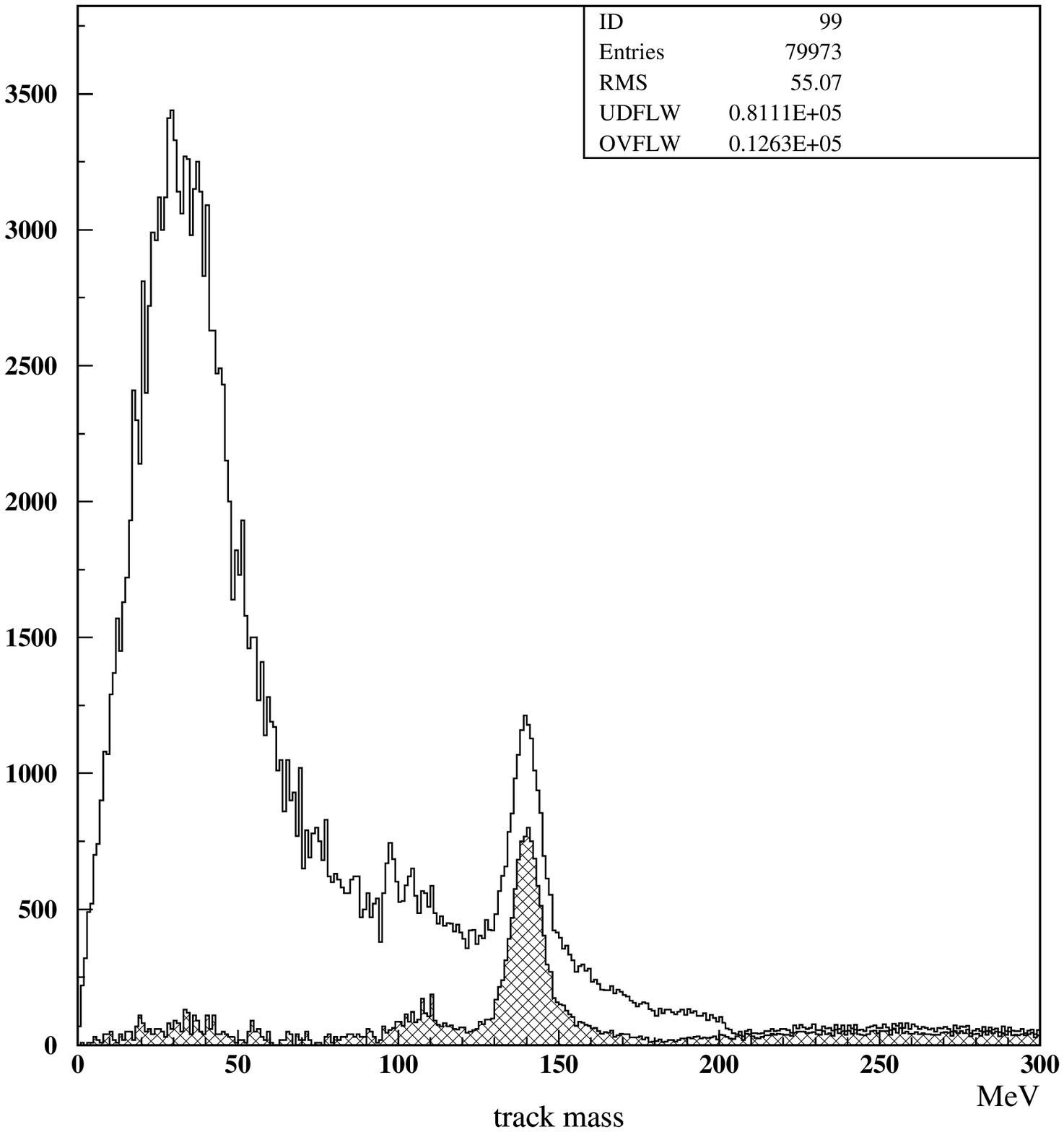,height=3.5in}
\psfig{figure=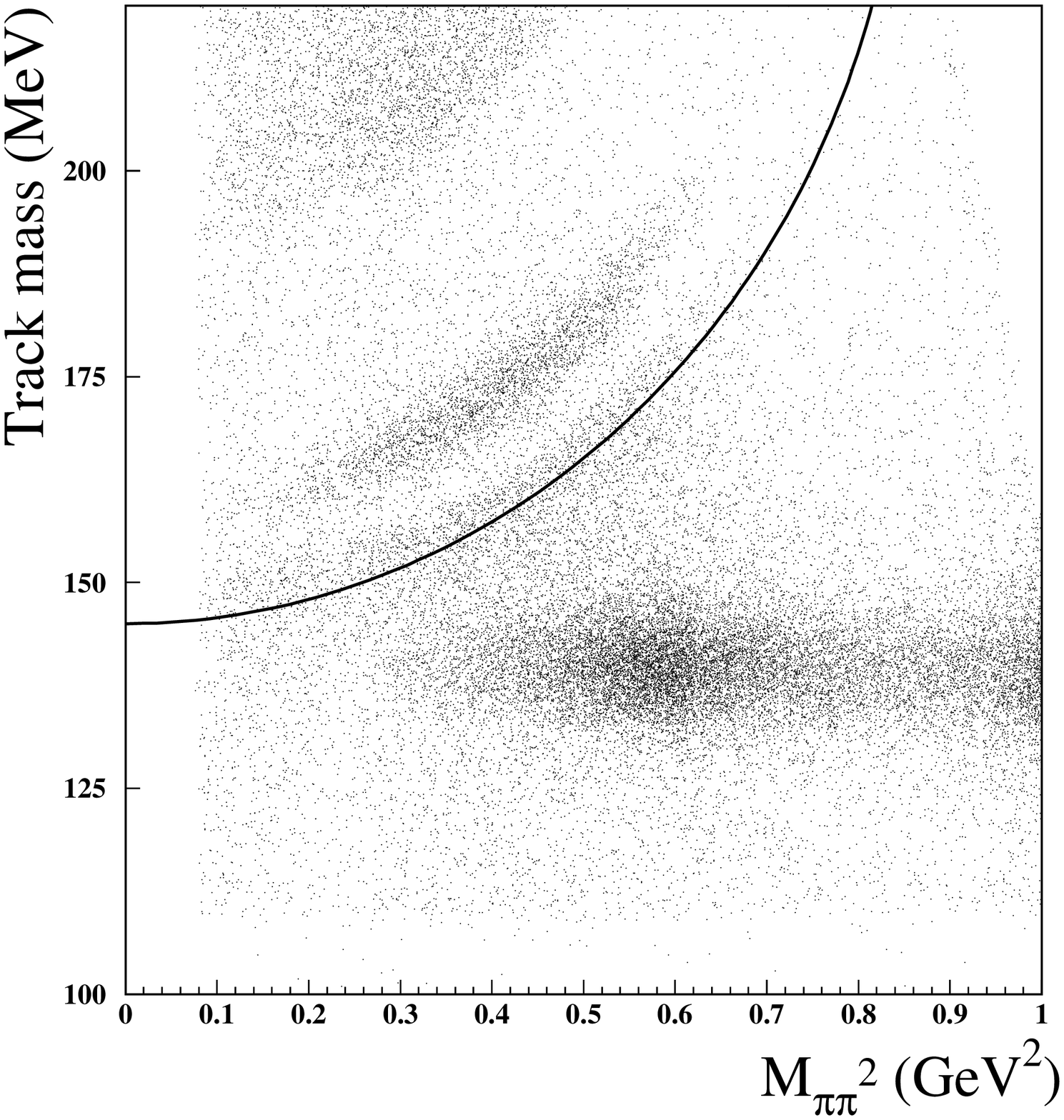,height=3.5in}
\caption{Plot on the left: track mass distribution before and after the likelihood selection. $\mmg$, $\ppg$ distributions are peaked at the physical mass, three pion events populate the region on the right of the $\ppg$ peak. Plot on the right: cut in the plane ($M_{\pi\pi}^{2}$, track mass) used to reject $\pi^{+}\pi^{-}\pi^{0}$ events.
\label{fig:trkmass}}
\end{center}
\end{figure}
This variable is calculated from the reconstructed momenta, $\vec{p}_{+}$, $\vec{p}_{-}$, applying 4-momentum conservation, under the hypothesis that the final state consists of two particles with the same mass and one photon. For each event class ($\mmg$, $\ppg$) the track mass distribution is peaked at the physical mass. $\pi^{+}\pi^{-}\pi^{0}$ events populate the region on the right of the $\ppg$ peak. 
\item {\it cut on the track mass}. $\mmg$ events are rejected by a cut at 120 MeV in the track mass. The discrimination between $\pi$ and $\mu$ using calorimeter information is not helpful since pions behave frequently like minimum ionizing particles. After this cut we find a contamination of $\mmg$ background smaller than $1\%$. 
$\pi^{+}\pi^{-}\pi^{0}$ events are rejected with a cut in the two-dimensional distribution of the track mass versus the two pion invariant mass squared, $M_{\pi\pi}^{2}$, shown in the right plot of Fig.~\ref{fig:trkmass}. Due to the large production cross section for these events ($BR(\phi \rightarrow \pi^{+}\pi^{-}\pi^{0})\sim 15\%$, $\sigma\sim500$ nb) and the similar kinematics, this background is hard to reject completely. The largest background contamination is expected at small $M_{\pi\pi}^{2}$ values ($M_{\pi\pi}^{2} < 0.4-0.5$ GeV$^{2}$). 
The efficiency of the track mass cut, as evaluated from MC, is $\sim 90\%$.
\item {\it definition of the angular acceptance}. The polar angle of the two pion system, $\theta_{\pi\pi}$, is calculated from the charged tracks. Two fiducial volumes for the analysis of $\ppg$ events are defined: $\theta_{\pi\pi} < 15^{o}$ or $\theta_{\pi\pi} > 165^{o}$ (Small Angle analysis) and $55^{o}< \theta_{\pi\pi} < 125^{o}$ (Large Angle analysis). In the following sections we discuss why these acceptance cuts are used. 
\end{itemize}
\subsection{Small angle analysis}
A challenging task in the analysis of $e^{+}e^{-} \rightarrow \ppg$ events is the suppression of Final State Radiation (FSR) events, where the photon is emitted by one of the pions. Since an initial state photon is emitted preferably at small polar angles, while the polar angle distribution of the photon from Final State Radiation follows the pion sin$^{2}\theta_{\pi}$ distribution, the signal to background ratio changes considerably with the polar angle of the two pion system, $\theta_{\pi\pi}$. Hence an adequate choice of the fiducial volume of $\theta_{\pi\pi}$ helps in the suppression of this background. 

The selection of events with $\theta_{\pi\pi}$ within a cone of $15^{o}$ around the beam pipe rejects almost completely FSR (from Monte Carlo the remaining contamination is expected to be smaller than $0.5\%$). Similarly the $\pi^{+}\pi^{-}\pi^{0}$ background is largely suppressed by these acceptance cuts. 

In the small angle analysis the ISR cross section at the $\rho$ resonance is very large, but the region $M_{\pi\pi}^{2} < 0.25$ GeV$^{2}$ is kinematically not accessible. 
\subsection{Large angle analysis}
In the large angle analysis the background conditions are worse: the FSR to ISR ratio can be of the order of $50\%$ at large values of the two pion invariant mass. The $\pi^{+}\pi^{-}\pi^{0}$ contamination is also much larger than in the small angle region. 

Howewer, the selection of the hadronic system at large angles has the merit that the full energy range of the two pion, $4m_{\pi}^{2} < M_{\pi\pi}^{2} < m_{\phi}^{2}$, can be explored. 
Since the photon emitted at large angles is detected by the electromagnetic calorimeter, more stringent kinematic constraints can be applied in this analysis. 

Near the $\pi\pi$ threshold, where the three pion contamination is larger, the direction of the photon is very well reconstructed from the charged tracks. The matching between this direction and the one reconstructed from the electromagnetic calorimeter rejects most of $\pi^{+}\pi^{-}\pi^{0}$ background. 
Since at the threshold the radiated photon is very energetic and almost back to back to any of the pions, the FSR contribution is also negligible. 
\section{Effective cross section}
We present here the result of the analysis of 22.6 pb$^{-1}$ collected in 2000. 265200 events are selected in the small angle analysis. The $\pi^{+}\pi^{-}\pi^{0}$ contamination is extremely small ($\sim$ 0.2$\%$). 46715 events are selected in the large angle analysis, the contamination of $\pi^{+}\pi^{-}\pi^{0}$ is about $4\%$ and is concentrated in the region $M_{\pi\pi}^{2} < 0.6$ GeV$^{2}$. In both cases the three pion background surviving the signal selection has been evaluated from Monte Carlo. 

From the experimental distributions in $M_{\pi\pi}^{2}$ at large and at small angles we obtain two effective cross sections by subtracting the number of $\pi^{+}\pi^{-}\pi^{0}$ background events from the observed number of events and normalizing to the integrated luminosity, as in the following:
\begin{equation}
\frac{d\sigma(e^{+}e^{-} \rightarrow \ppg)}{dM_{\pi\pi}^{2}} = \frac{dN_{obs}-dN_{bkg}}{dM_{\pi\pi}^{2}}\cdot\frac{1}{\int {\cal} L dt}
\label{eq:eff_cross}
\end{equation}
In the next sections we discuss the measurement of integrated luminosity at KLOE and show the comparison of the experimental effective cross sections with the corresponding Monte Carlo predictions. 
\subsection{Luminosity measurement}
The DA$\Phi$NE accelerator does not have luminosity monitors at small angle due to the existence of focusing quadrupole magnets very close to the interaction point. 
The luminosity is therefore measured using Large Angle Bhabhas ($55^{o} < \theta_{+,-} < 125^{o}$, $\sigma=425$nb). The number of LAB candidates are counted and normalized to the effective Bhabha cross section obtained from Monte Carlo. 

The precision of this measurement depends both on the understanding of experimental efficiencies and acceptances and on the theoretical knowledge of the process. 

The systematic errors arising from the LAB selection cuts are well below $1\%$. All the selection efficiencies concerning the LAB measurement (Trigger, EmC clusters, DC tracking) are above $98\%$ and well reproduced by the detector simulation. The background due to $\mmg$, $\ppg$ and $\pi^{+}\pi^{-}\pi^{0}$ is below $1\%$. 

KLOE uses two independent Bhabha event generators (the Berends/Kleiss generator~\cite{berends} modified for DA$\Phi$NE~\cite{graziano} and BABAYAGA~\cite{babayaga}). 

The very good agreement of the experimental distributions ($\theta_{+,-}$, $E_{+,-}$) with the event generators and a cross check with an independent luminosity counter based on $e^{+}e^{-} \rightarrow \gamma \gamma(\gamma)$ indicates a precision of better than $1\%$. 

More systematics checks (e.g. the effect of a varying beam energy and of a displaced beam interaction point) are underway.
\subsection{Comparison with the Monte Carlo}
\begin{figure}
\begin{center}
\psfig{figure=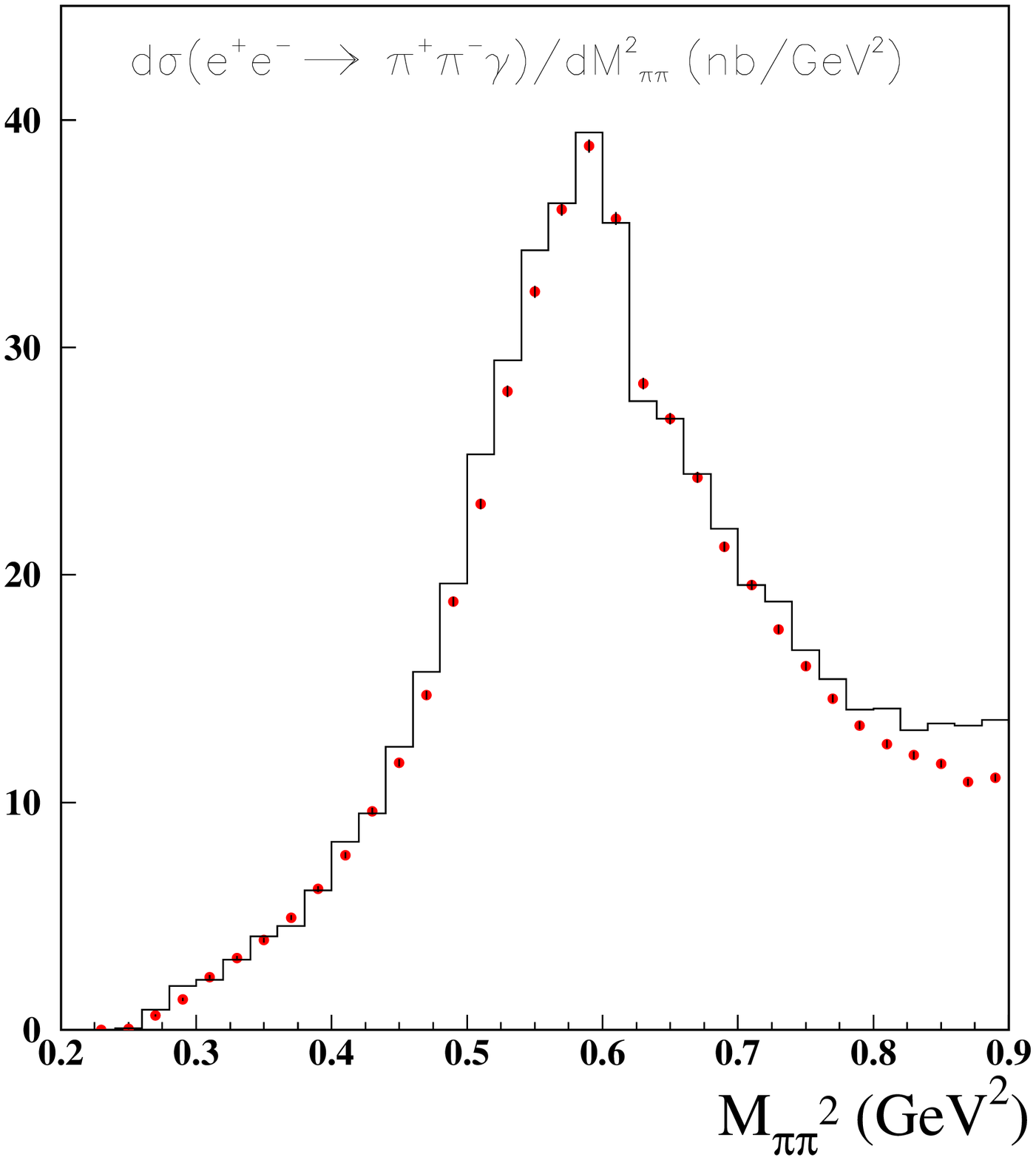,height=3.5in}
\psfig{figure=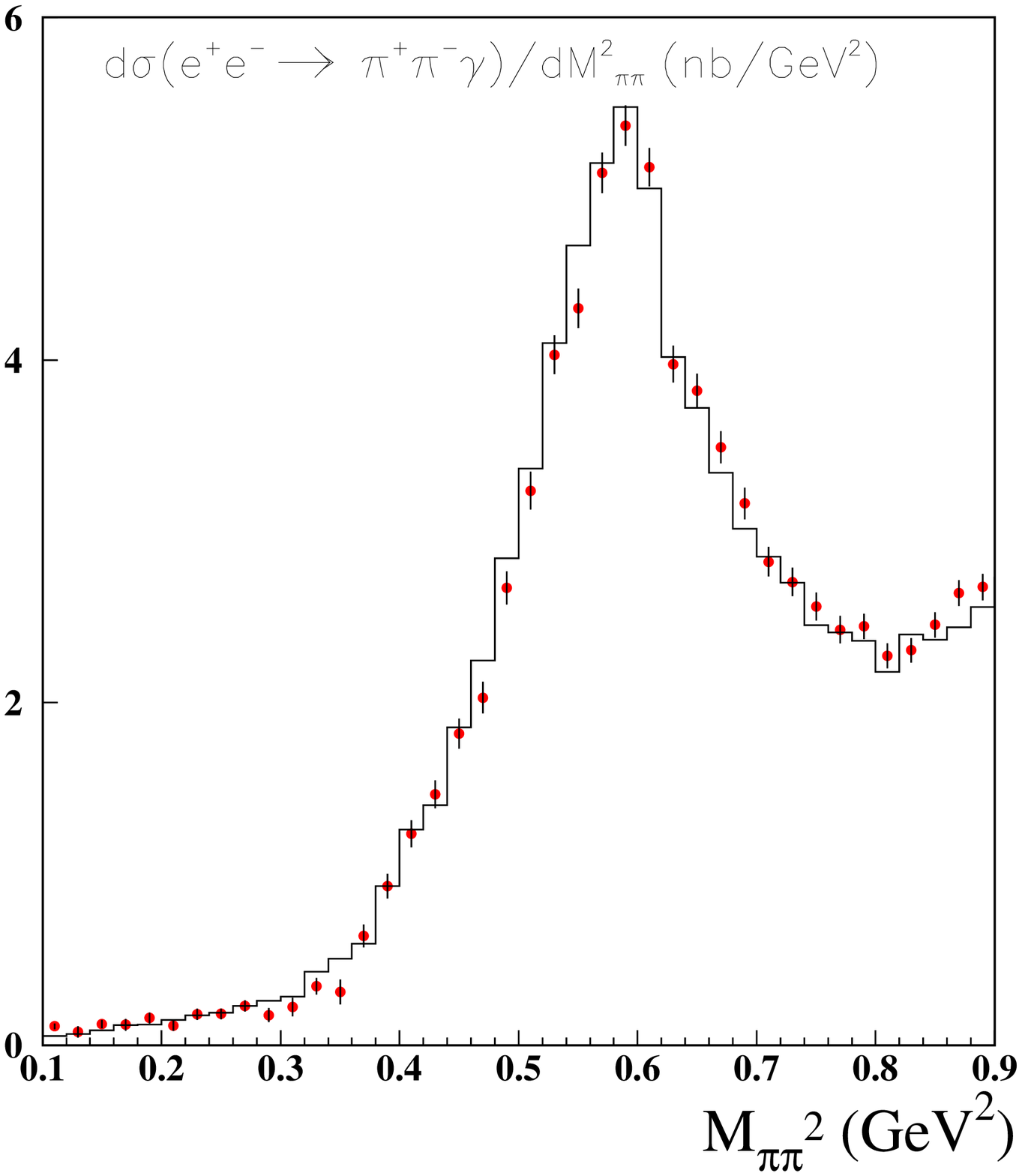,height=3.5in}
\caption{Effective cross sections at small (on the left) and at large photon angles (on the right). The solid line is the prediction of the $\ppg$ generator EVA. The sample corresponds to an integral luminosity of 22.6 pb$^{-1}$.
\label{fig:sigma}}
\end{center}
\end{figure}

We present in fig.~\ref{fig:sigma} the effective cross sections obtained from the analysis at small and at large angles, as defined in Eq.~\ref{eq:eff_cross}. The solid line shows the predictions of our $\ppg$ generator, EVA~\cite{bkm}, including the full detector simulation and the complete analysis procedure. 

The selection efficiency has been evaluated basically from data, as said in sec.~\ref{subsec:sig_sel}. 

In both analyses we obtain good agreement with the Monte Carlo. The effective cross section obtained at small angles with EVA has to be improved since the cross section (simulating only leading order and collinear radiation), diverges for $\theta_{\pi\pi} \rightarrow 0$. For a more realistic comparison the complete next to leading order generator is necessary. This generator is now available~\cite{rodrigo} and will be soon inserted in the KLOE detector simulation. 

At large angles most of the errors come from background subtraction. More work is in progress for a better understanding of the $\pi^{+}\pi^{-}\pi^{0}$ contamination.

%
\section{Summary and outlook}
We presented in this paper a preliminary measurement of $d\sigma(e^{+}e^{-} \rightarrow \ppg)/dM_{\pi\pi}^{2}$ for $M_{\pi\pi} < 1$ GeV$^{2}$ using the radiative return method. The data sample corresponds to an integrated luminosity of 22.6 pb$^{-1}$. Two different analyses have been performed, one at small angles ($\theta_{\pi\pi} < 15^{o}$ or $\theta_{\pi\pi} > 165^{o}$) and one at large angles ($55^{o} < \theta_{\pi\pi} < 125^{o}$). The $\rho$ resonance can be studied with high statistics in the analysis at small angles. The merit of this analysis is the very low contamination of $\pi^{+}\pi^{-}\pi^{0}$ background ($\sim 0.2\%$) and the small contribution of Final State Radiation (below $0.5\%$). The analysis at large angles allows to explore the hadronic cross section down to the $\pi\pi$ threshold. At low $M_{\pi\pi}^{2}$ values stringent constraints on the kinematics can been applied to reject $\pi^{+}\pi^{-}\pi^{0}$ background, while the contribution of FSR is expected to be negligible. Both analyses show good agreement with the $\rho$ parametrization of EVA. We conclude that the experimental understanding of efficiencies, background and luminosity are well under control. We further conclude that the analysis has demonstrated the great potential of the radiative return method to measure hadronic cross section. 

In order to improve the accuracy of $a_{\mu}$ and to be competitive with results coming from the CMD-2 experiment in Novosibirsk~\cite{cmd}, a final precision for this measurement below $1\%$ is needed. KLOE has collected 175 pb$^{-1}$ in 2001. With this data sample we can achieve an overall statistical accuracy on $\sigma (e^{+}e^{-} \rightarrow \pi^{+}\pi^{-})$ better than $0.3 \%$. Furthermore more refined systematics studies can be done with the larger statistics available. An improvement in the knowledge of the radiative corrections to the process $e^{+}e^{-} \rightarrow \pi^{+}\pi^{-}$  and on the luminosity measurement is also expected thanks to the collaboration with different theoretical groups~\cite{rodrigo}~\cite{jege}~\cite{babayaga}.  
\section*{Acknowledgments}
This work has been partially supported by ``European Community - Access to Research Infrastructure, contract number HPRI-CT 1999-00088''. 
\section*{References}


\begin{thebibliography}{99}
\bibitem{dafne}S. Guiducci {\it et al}, Proceedings of PAC99, New York, March 1999.
\bibitem{spagnolo} S. Spagnolo \Journal{{\em Eur. Phys. J.} C}{6}{637}{1999}.
\bibitem{bkm}S. Binner, J.H. K\"uhn, K. Melnikov, \Journal{\PLB}{459}{279}{1999}.
\bibitem{rodrigo}J.H. K\"uhn, G. Rodrigo, hep-ph/0204283, H. Czyz, J. H. K\"uhn, G. Rodrigo, hep-ph/0112184, G. Rodrigo, \Journal{{\em Acta Phys. Polon.} B}{32}{3833}{2001}.
\bibitem{jege} A. Hoefer, G. Gluza, F. Jegerlehner, {\it Pion Pair Production with Higher Order Radiative Corrections in Low Energy $e^{+}e^{-}$ collisions}, {\bf DESY-00-163} (2001).
\bibitem{pancheri} V.A. Khoze, M.I. Konchatnij, N.P. Merenkov, G. Pancheri, L. Trentadue, O.N. Shekhovyova, \Journal{{\em Eur. Phys. J.} C}{18}{481}{2001}.
\bibitem{jege2} F. Jegerlehner, {\it Hadronic Contributions to the Photon Vacuum Polarization and their role in Precision Physics}, {\bf DESY-01-028} (2001).
\bibitem{davier} M. Davier and A. H\"ocker, \Journal{\PLB}{435}{427}{1998}. 
\bibitem{berends} F.A. Berends, R. Kleiss, \Journal{\NPB}{228}{537}{1988}.
\bibitem{graziano} E. Drago, G. Venanzoni, {\it A Bhabha Generator for DA$\Phi$NE including radiative corrections and $\phi$ resonance}, {\bf INFN/AE97/48} (1997).
\bibitem{babayaga} C.M.C. Calame, C. Lunardini, G. Montagna, O. Nicrosini, F. Piccinini, \Journal{\NPB}{584}{459}{2000}.
\bibitem{cmd} R.R. Akhmetshin {\it et al.} [CMD-2 Collaboration], \Journal{\PLB}{527}{161}{2002}.
\end{thebibliography}
\end{document}